\documentclass[aps,prl,showpacs, twocolumn]{revtex4}

\usepackage{graphicx}
\usepackage{bm}
\usepackage{amsmath}

\providecommand{\bra}[1]{\langle #1 \rvert}
\providecommand{\ket}[1]{\lvert #1 \rangle}

\providecommand{\be}{\begin{equation}}
\providecommand{\ee}{\end{equation}}
\providecommand{\ba}{\begin{eqnarray}}
\providecommand{\ea}{\end{eqnarray}}

\usepackage{mathbbol}
\begin{document}

\title{Quantum information with modular variables}
 
\author{A. Ketterer$^1$, S. P. Walborn$^2$, A. Keller$^3$, T. Coudreau$^1$ and P. Milman$^1$}

\affiliation{$^{1}$Laboratoire Mat\'eriaux et Ph\'enom\`enes Quantiques, Universit\'e Paris Diderot, CNRS UMR 7162, 75013, Paris, France}
\affiliation{$^{2}$ Instituto de F\'{\i}sica, Universidade Federal do Rio de Janeiro. Caixa Postal 68528, 21941-972 Rio de Janeiro, RJ, Brazil}
\affiliation{$^{3}$Univ. Paris-Sud 11, Institut de Sciences Mol\'eculaires d'Orsay (CNRS), B\^{a}timent 350--Campus d'Orsay, 91405 Orsay Cedex, France}

\begin{abstract}
We introduce a novel strategy, based on the use of modular variables, to encode and deterministically process quantum information using  states described by continuous variables. Our formalism leads to a general recipe to adapt existing quantum information protocols, originally formulated for finite dimensional quantum systems, to infinite dimensional systems described by continuous variables. This is achieved by using non unitary and  non-gaussian operators, obtained from the superposition of gaussian gates,  together with  adaptative manipulations in  qubit systems defined in  infinite dimensional Hilbert spaces. We describe in details the realization of single and two qubit gates and briefly discuss their implementation in a quantum optical set-up. 
\end{abstract}
\vskip2pc 
 
\maketitle

\paragraph{Introduction} Quantum information protocols are usually formulated in terms of qubits that are  manipulated to realize computational and communication tasks that may overperform their classical analogs. Some of the successes of qubit based quantum algorithms, are Shor's factorization algorithm \cite{Shor}, Grover's search algorithm \cite{Grover1, Grover2}, quantum random walk \cite{Julia} among many others \cite{site}. On the side of quantum communication protocols, quantum key distribution devices \cite{Ekert, BB84} are commercialized  \cite{IDQuantique}. We can also mention the possibility of quantum teleportation \cite{Teleportation} or steering  \cite{EPRsteering1, EPRsteering2}. A natural question is whether quantum information protocols implementations are restricted to discrete finite systems. Can  the quantum properties of Hilbert spaces of infinite dimensions be also used to encode quantum information?  Apart from its fundamental importance, answering this question can have important practical consequences since, in many cases, entangled continuous variable states have the appealing advantage that they can be prepared deterministically  and detected with high efficiency \cite{Loock}.  In \cite{Braunstein, Furosawa} it was shown that quantum teleportation can be realized using gaussian states. In \cite{Grosshans1, Grosshans2} quantum cryptography protocols relying on states defined by continuous variables were devised. Universality for manipulation of continuous variables quantum states was defined in \cite{Loyd, Loock}, where the necessity of non-gaussian operations was identified. Based on such a set of operations, measurement based quantum computation was generalized from the discrete to the continuous realm \cite{MBQC}.  Nevertheless, even if some quantum communication protocols are successfully formulated for continuous and discrete variables, and some essential tools for quantum state manipulation were identified in both realms, the field of continuous variables quantum algorithms certainly does not have the same impact as its discrete counterpart \cite{Pati, Pati1}.  In \cite{Gottesman}, it was shown that it is possible to encode a qubit in a quantum oscillator using superpositions of infinitely squeezed states, and from this idea, some quantum error correcting codes analogous to discrete ones were formulated.

In spite of the vast literature on the subject, several open questions remain. In the present contribution, we address the following, which is perhaps  the most important one: is it possible to design a general, dimension independent, formalism for quantum information protocols? We answer positively to this question introducing a formalism relying on modular variables \cite{Aharanov}. This formalism has recently been used to show the possibility of continuously discretizing infinite dimensional Hilbert spaces \cite{Pierre}, and related ideas proved themselves useful to provide general conditions to test the Clauser-Horne-Shimony-Holt (CHSH)  \cite{CHSH, Andreas} and  the Legget-Garg \cite{LG, Rabl} inequalities. Our  positive answer to this question provides a recipe to bridge the worlds of continuous and discrete variables based quantum information. Consequently, it  allows for the direct adaptation of quantum algorithms and quantum logic gates originally conceived for qubit systems to continuous variables ones, and we provide an example of this in the present manuscript. Finally, our results open the perspective of comparing on a same ground quantum information protocols that exist in both realms, but that are, so far, apparently disconnected. 

This paper is organized as follows: we start by recalling the modular variables formalism, and then show how a generalized form of the non-unitary non gaussian operators defined in \cite{Pierre} can be used to manipulate quantum information. We then propose an architecture based on conditional operations for deterministic quantum information processing using the introduced operators. We conclude by illustrating our ideas using a quantum optical set-up. 

\paragraph{Continuous discretization and qubit identification:}
By defining dimensionless position, $\hat{\theta}=2\pi \hat x/l$, and momentum, $\hat k=l \hat p /\hbar$, it was shown in \cite{Aharanov} that these operators can be split  into an integer and a modular part:
\begin{equation}
 \hat{\theta} =2\pi \hat N + \hat{\bar \theta} \  \  \   \  \  \  \  \    \               \hat k = \hat M + \hat{\bar k},
\label{eq:DefModVar}
\end{equation}
where $\hat N$ and $\hat M$ have integer eigenvalues and $\hat{\bar \theta}$ and $\hat{\bar k}$ are the dimensionless modular position and momentum operators with eigenvalues in the intervals $[0,2\pi[$ and $[0,1[$, respectively. Since $[\hat{\bar \theta}, \hat{\bar k}]=0$ \cite{Aharanov},   we can define a common  eigenbasis $\{\ket{(\hat{\bar \theta},\hat{\bar k})}\}$, referred to as \textit{modular basis}. Arbitrary quantum states in either position or momentum representation can be expressed in this basis, as $\ket{\Psi}=\int_0^{2\pi}\int_0^1 d\bar \theta d\bar k\ g({\bar \theta},{\bar k})\ \ket{(\bar \theta,\bar k)}$, with a complex normalized coefficient function $g:[0,2\pi[\times[0,1[\rightarrow \mathbb C$ \cite{commentA}.

Labeling quantum states using bounded continuous variables, as is the case here, enables the definition of two disjoint sets of equal size for each one of the variables. For the sake of clarity and without loss of generality for the purposes of this manuscript, we now consider that only the domain of the variable $\bar \theta$ is split into two sets of equal size. Such a splitting can be done in infinitely many ways \cite{comment}, and in order to illustrate the principles of our ideas, we discuss in details its splitting into two subintervals $[0,\pi[$ and $[\pi,2\pi[$. Further on, this splitting enables the definition of a continuum of two-level systems spanned by the states $\{\ket{(\hat{\bar \theta},\hat{\bar k})},\ket{(\hat{\bar \theta}+\pi,\hat{\bar k})}\}$ with which we can express the general state $\ket\Psi$ as:
\begin{align}
\ket{\Psi}=\int_0^{\pi}d\bar \theta \int_0^1 d\bar k f(\bar \theta,\bar k) \ket{\tilde{\Psi}(\bar \theta,\bar k)},
\label{eqn:GeneralState}
\end{align}
where
\begin{align}
\ket{\tilde{\Psi}(\bar \theta,\bar k)}=&\cos{(\alpha(\bar \theta,\bar k)/2)} \ket{(\bar \theta,\bar k)} \nonumber \\
&+ \sin{(\alpha(\bar \theta,\bar k)/2)} e^{i\phi(\bar \theta,\bar k)} \ket{(\bar \theta+\pi,\bar k)},
\label{eqn:continuousqubit}
\end{align} 

with a $\pi$ periodic complex function 
$f(\bar \theta,\bar k)$ such that  $\int_0^{\pi}d\bar \theta \int_0^1 d\bar k |f(\bar \theta,\bar k)|^2 =1$ and two real functions, $\alpha(\bar \theta,\bar k) $ and $\phi(\bar \theta,\bar k)$, defined on $[0,\pi[\times[0,1[$. Equation~(\ref{eqn:GeneralState}) can be seen as  a weighted continuous superposition of  pure qubit states $\ket{\tilde{\Psi}(\bar\theta,\bar k)}$ for each subspace with fixed $\bar\theta$ and $\bar k$. We stress that, so far, no approximation has been made, and state (\ref{eqn:GeneralState}) is simply an alternative way of looking at an ${\it arbitrary}$ state expressed in a continuous basis. The advantage of such representation is that it identifies $\{\bar \theta, \bar k\}$ dependent qubits. Thus, it is now possible to define universal quantum operations and logic gates acting in such $\{\bar \theta, \bar k\}$ dependent two dimensional subspaces. ``Translation" of  quantum algorithms and protocols originally defined for qubits to the continuous variables realm can be done by implementing it independently in each one of the $\{\bar \theta, \bar k\}$ dependent two dimensional subspaces, an idea that has already been applied to determine the general conditions for violation of CHSH inequalities \cite{Andreas} and to implement a CV Grover search algorithm \cite{Andreas2} with a different approach from \cite{Pati1}. 

The above formulation  is  equivalent to defining qubits in CV in the following way: state $\ket{\bar 0}$ is associated to (\ref{eqn:GeneralState}) while state $\ket{\bar 1}$ is associated to the state for which {\it all} the $\{\bar \theta, \bar k\}$ dependent two dimensional states (expressed as  (\ref{eqn:continuousqubit})) are orthogonal.  A simple example of a CV qubit definition is obtained by taking  $\alpha(\bar \theta,\bar k)=0$ or $\alpha(\bar \theta,\bar k)=\pi/2 \ \ \forall \{\bar \theta, \bar k\}$ in (\ref{eqn:GeneralState}), which gives \cite{commentB}:
\begin{align}
\ket{\bar 0}&=\int_0^{\pi}d\bar \theta \int_0^1 d\bar k f(\bar \theta,\bar k) \ket{(\bar \theta,\bar k)}, \label{eqn:ContQubits0}  \\
\ket{\bar 1}&=\int_0^{\pi}d\bar \theta \int_0^1 d\bar k f(\bar \theta,\bar k) \ket{(\bar \theta+\pi,\bar k)}. 
\label{eqn:ContQubits1}
\end{align} 

We now discuss how such qubits can be manipulated with an analogous to the su(2) algebra based on the application of periodic observables with a continuous spectrum.  

\paragraph{Single qubit gates:}
In order to manipulate the  $\{\bar \theta, \bar k\}$ dependent qubit states and equivalently (\ref{eqn:ContQubits0}) and (\ref{eqn:ContQubits1}), we introduce a set of continuous periodic non-unitary and non gaussian hermitian operators  \cite{Pierre}:
\begin{eqnarray}
\hat \Gamma_{\alpha} = \int_0^{\pi} d\bar \theta \int_0^1 d\bar k \  \zeta ({\bar \theta},{\bar k}) \hat \sigma_{\alpha}({\bar \theta},{\bar k}),
\label{eq:Gammas}
\end{eqnarray}
where $\alpha=x,y,z$, $\hat \sigma_{\alpha}({\bar \theta},{\bar k})$ are $SU(2)$ generators defined in each subspace $\{\ket{\{{\bar \theta},{\bar k}\}},\ket{\{{\bar \theta}+\pi,{\bar k}\}}\}$ and $\zeta({\bar \theta},{\bar k})$ is a real weight function \cite{comment2}, so that operators (\ref{eq:Gammas}) are hermitian. Notice that operators (\ref{eq:Gammas}) act independently on each $\{\bar \theta, \bar k \}$ subspaces irrespectively of the exact form of  $\zeta({\bar \theta},{\bar k})$. For this reason, our formulation is dimension independent, the usual  qubit case being recovered for  $\zeta({\bar \theta},{\bar k})=\delta({\bar \theta-\bar \theta'},{\bar k-\bar k'}) $ and unitary $\hat \Gamma_{\alpha}$ operators (the Pauli matrices). A simple example of a $\hat \Gamma_{\alpha}$ operator with a continuous spectrum defined from the  position and momentum operators studied here, is obtained by using $\zeta ({\bar \theta},{\bar k}) =\cos{(\bar \theta-\bar k\pi)}$. This yields $\hat \Gamma_z=\cos{\hat \theta}$, {\it i.e.}, the cosine operator. In this example, operators $\hat \Gamma_{\alpha}$ are obtained by unitary transformations, equivalent to rotations of the  $\hat \Gamma_z$ operator.

We can now use the operators~(\ref{eq:Gammas}) to define single qubit operations on CV states. For example, applying the operator $\hat\Gamma_x$ ``flips" states $\ket{(\bar \theta,\bar k)}$ and $\ket{(\bar \theta+\pi,\bar k)}$. Thus, applying it to the state $\ket{\bar 0}=\int_0^{\pi}d\bar \theta \int_0^1 d\bar k f(\bar \theta,\bar k) \ket{(\bar \theta,\bar k)}$ yields the equivalent to a spin flip operation on an ordinary qubit state:
\begin{align}
\hat\Gamma_x \ket{\bar 0}=\int_0^{\pi}d\bar \theta \int_0^1 d\bar k f(\bar \theta,\bar k) \zeta (\bar \theta,\bar k) \ket{(\bar \theta+\pi,\bar k)}=\ket{\bar 1'},
\label{eqn:Flip}
\end{align}
where, due to the non-unitarity of the $\hat \Gamma_x$ operator, the weight function  after the application of $\hat \Gamma_x$ changed from $f(\bar \theta,\bar k)$ to $ f(\bar \theta,\bar k) \zeta(\bar \theta,\bar k)$, justifying the definition of a  state $\ket{\bar 1'}$.

Nevertheless, state $\ket{\bar 1'}$ in (\ref{eqn:Flip}) is uniquely determined and satisfies $\langle \bar 1'|\bar 0\rangle=0$. Before discussing issues related to the non unitarity of the defined operators, we discuss arbitrary single qubit rotations. 

From  the previous observations, we have that arbitrary single qubit-like rotations, defined here as the non unitary operations $\hat U(\hat\Gamma_{\alpha}, \beta)$, can be realized  by superposing the operators~(\ref{eq:Gammas}) with a modulated identity operator $\mathbb 1_{\zeta}=\int_0^{\pi}d\bar \theta \int_0^1 d\bar k \zeta(\bar \theta,\bar k) \mathbb 1(\bar \theta,\bar k)$, with $\mathbb 1(\bar \theta,\bar k)=\ket{(\bar \theta, \bar k)}\bra{(\bar \theta, \bar k)}$, analogously to the usual qubit rotations: 
\begin{align}\label{eq:QubitRot}
\hat U(\hat\Gamma_{\alpha}, \beta) &\equiv \int_0^{\pi}  d\bar \theta \int_0^1 d \bar k \zeta(\bar \theta, \bar k) e^{i\beta {\hat \sigma_{\alpha}}(\bar \theta, \bar k)} \nonumber \\
&= \cos\beta \mathbb{1_{\zeta}}+ i\sin{\beta} \hat\Gamma_{\alpha}.
\end{align}
In order to analyze their action on states  (\ref{eqn:ContQubits0}) and (\ref{eqn:ContQubits1}), we study the specific example of  $\alpha=x$, leading to:
\begin{eqnarray}\label{Uqubit1}
&&\hat U(\hat\Gamma_{x}, \beta)\ket{\bar 0}=\cos \beta \ket{\bar 0'}+i\sin\beta \ket{\bar 1'}, \nonumber \\
&&\hat U(\hat\Gamma_{x}, \beta)\ket{\bar 1}=\cos \beta \ket{\bar 1'}+i\sin\beta \ket{\bar 0'},
\end{eqnarray}
where, analogously to (\ref{eqn:Flip}), $\ket{\bar 0'}$ is defined as 
\begin{align}
\hat\Gamma_x \ket{\bar 1}=\int_0^{\pi}d\bar \theta \int_0^1 d\bar k f(\bar \theta,\bar k) \zeta (\bar \theta,\bar k) \ket{(\bar \theta,\bar k)}=\ket{\bar 0'}.
\label{eqn:Flip2}
\end{align}
Thus, operators $\hat U(\hat\Gamma_{\alpha}, \beta)$ have the analogous effect of  a rotation in a qubit system when acting on states (\ref{eqn:ContQubits0}) and (\ref{eqn:ContQubits1}) as well, with the difference that they output non normalized states, a consequence of the non unitarity of operators $\hat U(\hat\Gamma_{\alpha}, \beta)$. 

\paragraph{Two-qubit gates} 
We now describe how to construct entangling gates,  as the controlled-NOT (CNOT) gate, so as to complete a set of universal quantum gates \cite{Chuang}.  We proceed  analogously to single qubit gates, but now  taking the sum of the tensor product of two $\hat\Gamma$-operators, respectively, as for instance:
\begin{widetext}
\begin{align}\label{eq:MultiGamma}
&\hat \Gamma_{\alpha_1}\otimes \hat \Gamma_{\beta_1} + \hat \Gamma_{\alpha_2}\otimes \hat \Gamma_{\beta_2} =&\\
&\int_0^{\pi}\int_0^{1} d\bar{\boldsymbol\theta}d\bar{\boldsymbol k}\left[\zeta_{a_1}({\bar \theta}_a,{\bar k}_a) \zeta_{b_1}({\bar \theta}_b,{\bar k}_b) \hat \sigma_{\alpha_1}(\hat{\bar \theta}_a,\hat{\bar k}_a)\otimes \hat \sigma_{\beta_1}(\hat{\bar \theta}_b,\hat{\bar k}_b)
+\zeta_{a_2}({\bar \theta}_a,{\bar k}_a) \zeta_{b_2}({ \bar \theta}_b,{\bar k}_b) \hat \sigma_{\alpha_2}(\hat{\bar \theta}_a,\hat{\bar k}_a)\otimes \hat \sigma_{\beta_2}(\hat{\bar \theta}_b,\hat{\bar k}_b) \right],&\nonumber
\end{align}
\end{widetext}
where $d\bar{\boldsymbol\theta}=d\bar \theta_1 d\bar \theta_2$ and $d\bar{\boldsymbol k}=d\bar k_1 d\bar k_2$. Again, we can define entangling gates acting independently in $\{\bar \theta_a, \bar \theta_b, \bar k_a, \bar k_b\}$ dependent subspaces, which are analogous to two-qubit systems, completing the set of universal quantum gates. Equivalently, (\ref{eq:MultiGamma}) leads to the construction of entangling gates and a universal set of operations acting on states (\ref{eqn:ContQubits0}) and (\ref{eqn:ContQubits1}).

\paragraph{Ancilla-driven  (positive operator valued measures) implementation:}
We now discuss on the non unitarity of the operators  used to implement the single and two qubit like quantum operations as well as on an architecture of deterministic quantum computation for systems in Hilbert spaces of arbitrary dimensions. Observables as (\ref{eq:Gammas}) can be considered as the elements of a two-valued generalized measurement. For $\hat \Gamma_z=\cos{\hat \theta}$, for instance, the measurement can be completed by adding another observable, $\hat \Gamma'_z=\sin{\hat \theta}$. These observables fulfill the completeness relation $\hat \Gamma_z\hat \Gamma_z+\hat \Gamma_z'\hat \Gamma_z'=\mathbb 1$, where the terms $\hat \Gamma_z\hat \Gamma_z$ and $\hat \Gamma_z'\hat \Gamma_z'$ are positive operator valued measurement (POVM). In this sense, one could argue that the operations defined above, allowing for the qubit-like manipulation of states described by continuous variables, are probabilistic. Nevertheless, this is not the case, since one can easily notice that observable $\hat \Gamma'_z$, in the discussed example, is analogous to $\Gamma_z$ {\it i.e.}: it manipulates each $\{\bar \theta, \bar k \}$ dependent two dimensional subspace independently in the same way as $\Gamma_z$, but with a different weight function $\zeta'(\bar \theta, \bar k)$. Even if the output state obtained from the action of $\Gamma_z$ is different from the one obtained from the action of $\Gamma'_z$, they can both be connected and transformed into each other by state independent operations if one knows which one of the  operators was applied. Thus, one can devise a strategy inspired of ancilla-driven quantum computation \cite{AncillaDrivenQC, POVMQC}, which is nothing but considering the extension of the two-valued POVM to a higher dimensional subspace including the ancilla qubits as well. Ancillae will provide information on which operation was realized. They do not represent a limitation of our protocol.

\begin{figure}[t]
\includegraphics[scale=0.4]{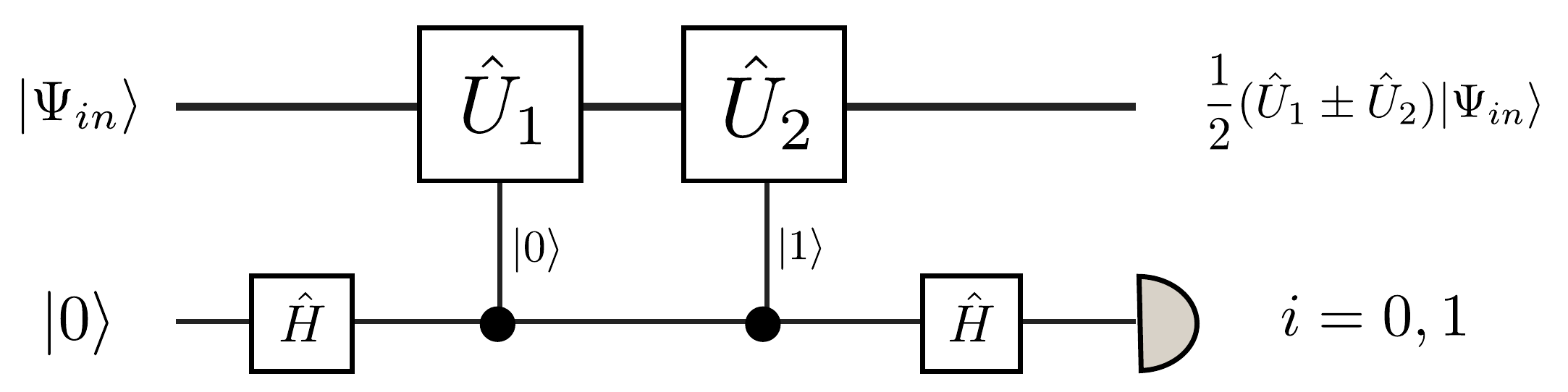}
\caption{Ancilla-driven continuous variable quantum circuit for the implementation of single qubit operations on the encoded CV qubit states. The initial system-ancilla state $\ket{\Psi_{in}} \ket{0}$ is transformed after the sequence of operations  described in the main text and represented in the circuit onto the final state $\hat H \hat{CU}_1\hat{CU}_2 \hat H \ket{\Psi_{in}}\ket 0$. The CV state is conditioned to the ancilla's.}
\label{fig:AncillaCircuit}
\end{figure}

The deterministic protocol to realize single qubit-like rotations consists of the following (see Fig. \ref{fig:AncillaCircuit}). The total  state of the system is described by an  arbitrary continuous input state $\ket\Psi$ and a quantum two-dimensional ancilla system with basis states $\ket{i}$ $i=1,2$ initially prepared in state $\ket 0$. The ancilla is then put  in a superposition state, and  performs controlled unitary operations, $\hat U_1$ and $\hat U_2$, on $\ket\Psi$. Finally, the ancilla is transformed as  $\ket {0}\rightarrow \frac{1}{\sqrt 2} (\ket 0 + \ket 1)$, $\ket {1}\rightarrow \frac{1}{\sqrt 2} (\ket 0 - \ket 1)$. The outcome of this sequence is such that:
\begin{align}
\ket{\Psi_i}=\frac{1}{2}(\hat U_1 +(-1)^i\hat U_2) \ket{\Psi_{in}} \hspace{0.5cm}i=0,1\label{eq:Circuit2}
\end{align}

The unitary operations $\hat U_1$ and $\hat U_2$ can be chosen so as to lead to operators Eq.~(\ref{eq:Gammas}). For example, by choosing $\hat U_1=e^{i\pi \hat k^2/2}e^{i \hat k}e^{-i\pi \hat k^2/2}$, $\hat U_2=\hat U_1^{\dagger}$, we obtain $\hat\Gamma_x\ket{\Psi_{in}}$ with $\zeta(\bar\theta, \bar k)=\cos{(\bar\theta-\pi \bar k)}$, if the  ancilla is in state  $\ket{0}$ and $\hat\Gamma_x'\ket{\Psi_{in}}$ with a different weight function $\zeta'(\bar \theta,\bar k)$ otherwise. By choosing
\begin{align}\label{US}
\hat U_k&=\frac{1}{\sqrt 2}  \left[ \cos(\alpha) \mathbb{1_{\zeta}}+ i\sin{(\alpha)}\hat\Gamma_{\hat n} \right.\\ \nonumber
&\left. +(-1)^{(k+1)}\cos(\alpha-\frac{\pi}{2}) \mathbb{1_{\zeta'}}+ (-1)^{(k+1)}i\sin{(\alpha-\frac{\pi}{2})}\hat\Gamma_{\hat n}'\right], 
\end{align} 
with $k=1,2$ and $\zeta^2+\zeta'^2=1$, arbitrary single qubit rotations (\ref{eq:QubitRot}) can be implemented, with  $\hat U(\hat\Gamma_x,\alpha)\ket{\Psi_{in}}$ if the ancilla is in state $\ket{0}$ or $\hat U(\hat\Gamma'_x,\alpha-\frac{\pi}{2})\ket{\Psi_{in}}$ if it is in state $\ket{1}$. 

Using non-unitary operations, and consequently, POVMs to manipulate CV quantum information is by no means a drawback and can lead to deterministic quantum information processing. Since different transformations are univocally associated to different (orthogonal) ancilla states, they can also be locally corrected through ancilla dependent operations. Such operations can simply be using different definitions of a qubit according to the ancilla state.  In the case described by Eq. (\ref{US}) this would correspond to interchanging the meaning of $\ket{\bar 0'}$ and $\ket{\bar 1'}$. 

\begin{figure}
\includegraphics[width=0.475\textwidth]{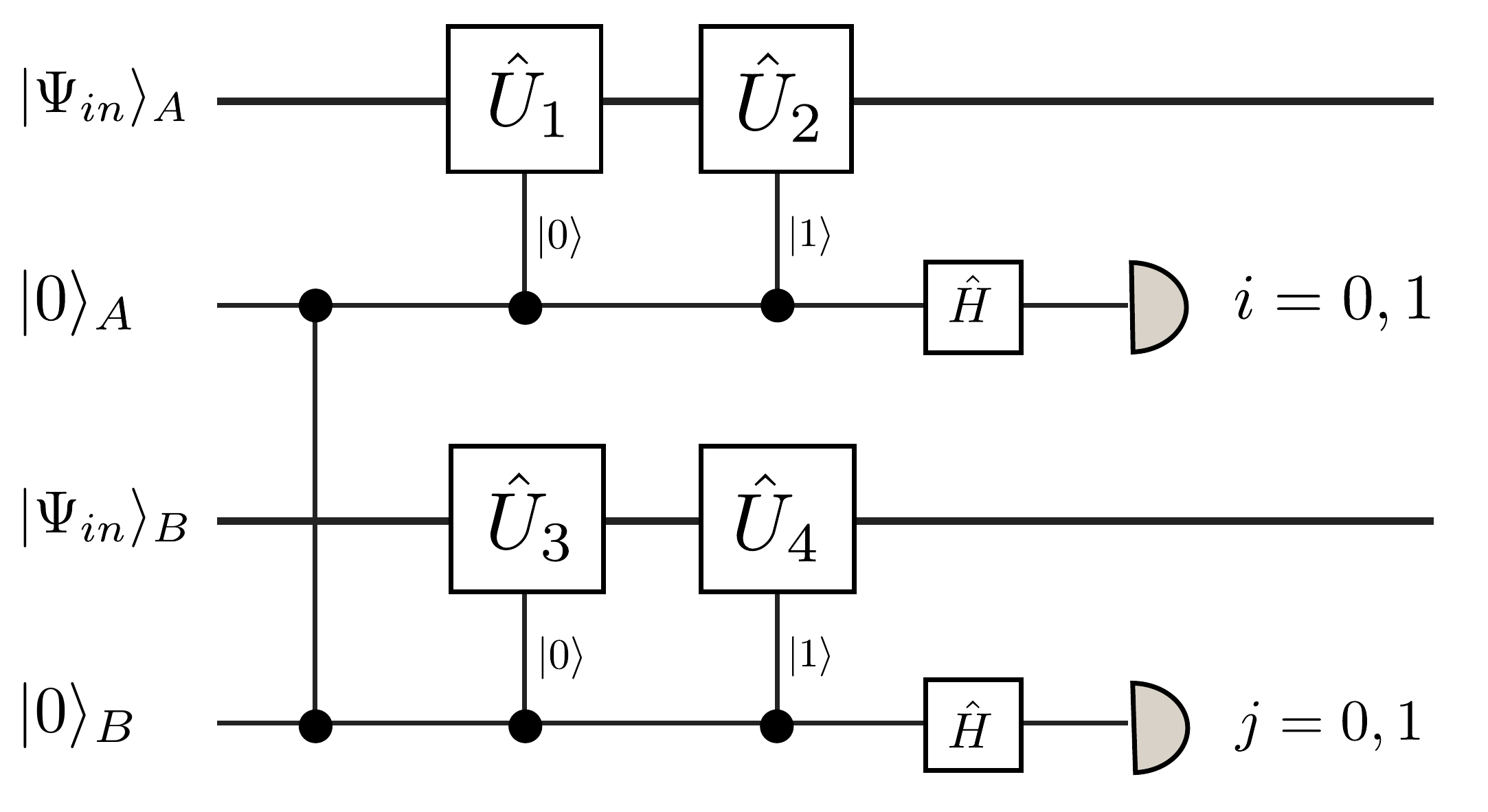}
\caption{Ancilla-driven continuous variable quantum circuit for the implementation of two-qubit operations on the encoded CV qubit states. The initial system-ancilla state $\ket{\Psi_{in}}_A \ket{0}_A\ket{\Psi_{in}}_B \ket{0}_B$ is mapped, via Hadamard operations on the ancilla and controlled unitary operations on the CV system, onto the final state $(\hat H\otimes \hat H) (\hat{CU}_1\otimes\hat{CU}_2)(\hat{CU}_3\otimes\hat{CU}_4) \hat E_{AB}  \ket{\Psi_{in}}\ket 0$. Different  ancilla states are associated to different  results  (\ref{eq:Circuit2}). }
\label{fig:AncillaCircuit2}
\end{figure}
Controlled two qubit operations via ancilla-driven quantum computation can be realized as shown in the circuit of  Fig. (\ref{fig:AncillaCircuit2}). Therein, the two ancilla qubits are first entanged, and then perform controlled unitary operations on the CV states of system A and B, respectively. Analogously to the single qubit case, different  two-ancillae states are associated to different final CV states:
\begin{align}\label{2qbit}
&\ket{\Psi_{ij}}=\frac{1}{4}((-1)^i\hat U_1 \otimes \hat U_2+(-1)^j\hat U_3 \otimes \hat U_4) \ket{\Psi_{in}} 
\hspace{0.25cm}i,j=0, 1. \nonumber \\
\end{align}
By choosing the appropriate unitary transformations one can implement different entangling operations on the CV qubit states (\ref{eqn:ContQubits0}) and (\ref{eqn:ContQubits1}), completing the set of universal quantum gates. Again,  from (\ref{2qbit}), the different entangling operations associated to each one of the two ancilla states can be mapped into a common deterministic operation by local correction of the operations or by redefinition of the encoding.

\paragraph{Transverse degree of freedom of photons:} The introduced formalism can be implemented in a number of experimental systems relying on ancillae to manipulate CVs, as the field's quadrature, the electromagnetic field in a cavity and the continuous degrees of freedom of a photon. We briefly describe how the interplay between ancillae and CV can be realized in the latter.  The transverse degrees of freedom of a photon are described by CV and are analogous to the field's quadrature \cite{Sergienko}. Different quantum states $\ket{\Psi}$ can be  engineered using spatial light modulators (SLM) \cite{PNAS}. On the other hand, the photon polarization or the propagation mode serve as a natural ancillae systems for each spatial CV system \cite{interferometer1, interferometer2}. Furthermore,  the polarization controlled CV unitary operations used in Fig. \ref{fig:AncillaCircuit} and \ref{fig:AncillaCircuit2} can be realized using SLMs \cite{Tasca}. 

Finally, the entangling operation on the ancilla states of systems A and B (see Fig. \ref{fig:AncillaCircuit}) can be realized using photons generated by spontaneous parametric downconversion. In this  way, an entangled state is created and can be used to implement the two or many qubit operations on the encoded CV qubits \cite{Zeilinger}.

\paragraph{Discussion and conclusion} We presented a formalism enabling the encoding of discretized information in arbitrary continuous variables states. Our results are obtained through the manipulation of the modular variables formalism, combined to the notion of ancilla-driven quantum computation and POVMs. As a consequence, our modular variable quantum computation protocol is based on the application of non-gaussian operations obtained through the superposition of gaussian ones. Our protocol dramatically simplifies the experimental requirements for encoding a qubit in quantum-harmonic oscillators \cite{Gottesman} or in $d\geq 2$ dimensional Hilbert spaces. It also allows for a direct translation between protocols defined for discrete variables to continuous ones. Last but not least, the universality of our approach suggests that using it to describe the existing quantum information protocols for continuous variables  may show that the latter, in fact, rely on the manipulation of quantum binary information.


\begin{thebibliography}{99}
\bibitem{commentA}  It is clear that such a decomposition is also possible for any pair of conjugated observables obtained from (\ref{eq:DefModVar}) by partial Fourier transform.
\bibitem{comment} In a more general treatment, qubits could also be defined using a different decomposition of the interval $[0,2\pi[$. The choice of decomposition can be determined, for instance, by experimental possibilities and constraints.
\bibitem{commentB} The function $f(\bar \theta, \bar k)$ used to define the qubit can be any. It can, for instance, be determined by experimental possibilities and constraints.
\bibitem{comment2} Notice that we have the freedom of choice with functions $\zeta_i(\bar \theta_i, \bar k_i)$. It can be any function such that its absolute value is $\pi$ periodic and continuous at the edges of $\pi$ intervals.
\end{thebibliography}
\end{document}